\documentclass[usenatbib]{mnras}
\usepackage{graphicx}
\usepackage{amsmath}
\usepackage{amssymb}
\usepackage{color}
\usepackage{url}

\usepackage{threeparttable}

\newcommand\lsim{\mathrel{\rlap{\lower4pt\hbox{\hskip1pt$\sim$}}
        \raise1pt\hbox{$<$}}}
\newcommand\gsim{\mathrel{\rlap{\lower4pt\hbox{\hskip1pt$\sim$}}
        \raise1pt\hbox{$>$}}}
\newcommand{\lya}{\ifmmode\mathrm{Ly}\alpha\else{}Ly$\alpha$\fi}
\newcommand{\lyb}{\ifmmode\mathrm{Ly}\beta\else{}Ly$\beta$\fi}
\newcommand{\igm}{\ifmmode\mathrm{IGM}\else{}IGM\fi}
\newcommand{\lae}{\ifmmode\mathrm{LAE}\else{}LAE\fi}
\newcommand{\h}{\ifmmode\mathrm{H}\else{}H\fi}
\newcommand{\hi}{\ifmmode\mathrm{H{\scriptscriptstyle I}}\else{}H{\scriptsize I}\fi}
\newcommand{\hii}{\ifmmode\mathrm{H{\scriptscriptstyle II}}\else{}H{\scriptsize II}\fi}
\newcommand{\cmb}{\ifmmode\mathrm{CMB}\else{}CMB\fi}
\newcommand{\qso}{\ifmmode\mathrm{QSO}\else{}QSO\fi}
\newcommand{\eor}{\ifmmode\mathrm{EoR}\else{}EoR\fi}
\newcommand{\heii}{\ifmmode\mathrm{He\,{\scriptscriptstyle II}}\else{}He\,{\scriptsize II}\fi}
\newcommand{\heiii}{\ifmmode\mathrm{He\,{\scriptscriptstyle III}}\else{}He\,{\scriptsize III}\fi}
\newcommand{\ciii}{\ifmmode\mathrm{C\,{\scriptscriptstyle III]}}\else{}C\,{\scriptsize III]}\fi}
\newcommand{\oiii}{\ifmmode\mathrm{O\,{\scriptscriptstyle III}}\else{}O\,{\scriptsize III}\fi}
\newcommand{\aliii}{\ifmmode\mathrm{Al\,{\scriptscriptstyle III}}\else{}Al\,{\scriptsize III}\fi}
\newcommand{\mgii}{\ifmmode\mathrm{Mg\,{\scriptscriptstyle II}}\else{}Mg\,{\scriptsize II}\fi}
\newcommand{\fe}{\ifmmode\mathrm{Fe}\else{}Fe\fi}
\newcommand{\nv}{\ifmmode\mathrm{N\,{\scriptscriptstyle V}}\else{}N\,{\scriptsize V}\fi}
\newcommand{\niv}{\ifmmode\mathrm{N\,{\scriptscriptstyle IV]}}\else{}N\,{\scriptsize IV]}\fi}
\newcommand{\cii}{\ifmmode\mathrm{C\,{\scriptscriptstyle II}}\else{}C\,{\scriptsize II}\fi}
\newcommand{\civ}{\ifmmode\mathrm{C\,{\scriptscriptstyle IV}}\else{}C\,{\scriptsize IV}\fi}
\newcommand{\siv}{\ifmmode\mathrm{Si\,{\scriptscriptstyle IV}}\else{}Si\,{\scriptsize IV}\fi}
\newcommand{\siii}{\ifmmode\mathrm{Si\,{\scriptscriptstyle II}}\else{}Si\,{\scriptsize II}\fi}
\newcommand{\siiii}{\ifmmode\mathrm{Si\,{\scriptscriptstyle III]}}\else{}Si\,{\scriptsize III]}\fi}
\newcommand{\ovi}{\ifmmode\mathrm{O\,{\scriptscriptstyle VI}}\else{}O\,{\scriptsize VI}\fi}
\newcommand{\sioiv}{\ifmmode\mathrm{Si\,{\scriptscriptstyle IV}\,\plus O\,{\scriptscriptstyle IV]}}\else{}Si\,{\scriptsize IV}\,+O\,{\scriptsize IV]}\fi}

\newcommand{\cmmc}{\textsc{\small 21CMMC}}
\newcommand{\cmfst}{\textsc{\small 21CMFAST}}

\pdfoutput=1
\title[LOFAR astrophysics]{Interpreting LOFAR 21-cm signal upper limits at $z \approx 9.1$ in the context of high-$z$ galaxy and reionisation observations}

\author[B. Greig et al.] {Bradley~Greig$^{1,2}$\thanks{E-mail:~greigb@unimelb.edu.au}, Andrei~Mesinger$^{3}$, L\'eon V. E. Koopmans$^{4}$, Benedetta Ciardi$^{5}$, \newauthor Garrelt Mellema$^{6}$, Saleem Zaroubi$^{7,8,4}$, Sambit K. Giri$^9$, Raghunath Ghara$^{7,8}$, \newauthor Abhik Ghosh$^{10}$, Ilian T. Iliev$^{11}$, Florent G. Mertens$^{4,12}$, Rajesh Mondal$^{6, 11}$,  \newauthor Andr\'e R. Offringa$^{4,13}$, Vishambhar N. Pandey$^{4,13}$ \\
$^1$ARC Centre of Excellence for All-Sky Astrophysics in 3 Dimensions (ASTRO 3D), University of Melbourne, VIC 3010, Australia \\
$^2$School of Physics, University of Melbourne, Parkville, VIC 3010, Australia \\
$^3$Scuola Normale Superiore, Piazza dei Cavalieri 7, I-56126 Pisa, Italy \\
$^4$Kapteyn Astronomical Institute, University of Groningen, PO Box 800, 9700 AV Groningen, The Netherlands \\
$^5$Max-Planck Institute for Astrophysics, Karl-Schwarzschild-Stra{\ss}e 1, D-85748 Garching, Germany \\
$^6$The Oskar Klein Centre, Department of Astronomy, Stockholm University, AlbaNova, SE-10691 Stockholm, Sweden \\
$^7$Department of Natural Sciences, The Open University of Israel, 1 University Road, PO Box 808, Ra'anana 4353701, Israel \\
$^8$Department of Physics, Technion, Haifa 32000, Israel \\
$^9$Institute for Computational Science, University of Zurich, Winterthurerstrasse 190, 8057 Zurich, Switzerland \\
$^{10}$Department of Physics, Banwarilal Bhalotia College, GT Rd, Ushagram, Asansol, West Bengal 713303, India \\
$^{11}$Astronomy Centre, Department of Physics and Astronomy, Pevensey II Building, University of Sussex, Brighton BN1 9QH, UK \\
$^{12}$LERMA, Observatoire de Paris, PSL Research University, CNRS, Sorbonne Universite, F-75014 Paris, France \\
$^{13}$ASTRON -- the Netherlands Institute for Radio Astronomy, Oude Hoogeveensedijk 4, 7991 PD Dwingeloo, The Netherlands 
}

\begin{document}
\maketitle \begin{abstract}
\noindent
Using the latest upper limits on the 21-cm power spectrum at $z\approx9.1$ from the Low Frequency Array (LOFAR), we explore regions of parameter space which are inconsistent with the data. We use \cmmc{}, a Monte Carlo Markov Chain sampler of \cmfst{} which directly forward models the 3D cosmic 21-cm signal in a fully Bayesian framework. We use the astrophysical parameterisation from \cmfst{}, which includes mass-dependent star formation rates and ionising escape fractions as well as soft-band X-ray luminosities to place limits on the properties of the high-$z$ galaxies. Further, we connect the disfavoured regions of parameter space with existing observational constraints on the Epoch of Reionisation such as ultra-violet (UV) luminosity functions, background UV photoionisation rate, intergalactic medium (IGM) neutral fraction and the electron scattering optical depth. We find that all models exceeding the 21-cm signal limits set by LOFAR at $z\approx9.1$ are excluded at $\gtrsim2\sigma$ by other probes. Finally, we place limits on the IGM spin temperature from LOFAR, disfavouring at 95 per cent confidence spin temperatures below $\sim2.6$~K across an IGM neutral fraction range of $0.15 \lesssim \bar{x}_{\hi{}} \lesssim 0.6$. Note, these limits are only obtained from 141~hrs of data in a single redshift bin. With tighter upper limits, across multiple redshift bins expected in the near future from LOFAR, more viable models will be ruled out. Our approach demonstrates the potential of forward modelling tools such as \cmmc{} in combining 21-cm observations with other high-$z$ probes to constrain the astrophysics of galaxies.
\end{abstract} 

\section{Introduction}

The Epoch of Reionisation (EoR) corresponds to the final major baryonic phase change of the Universe. This occurs when the once cold, neutral hydrogen that permeated the early Universe following recombination is ionised by the cumulative ionising radiation from astrophysical sources (e.g. stars and galaxies). Observing this phase transition is vitally important, as it reveals insights into the nature, growth and structure of the first astrophysical sources. These properties can be indirectly inferred through the imprint of their radiation on the intergalactic medium (IGM).

The ubiquity of neutral hydrogen in the early Universe should allow us to detect the imprint of the EoR through the 21-cm hyperfine transition \citep[see e.g.][]{Furlanetto:2006p209,Morales:2010p1274,Pritchard:2012p2958,Zaroubi:2013p2976,Barkana:2016}. This 21-cm signal is sensitive both to the thermal and ionisation state of the neutral hydrogen in the IGM. Further, being a line transition, the spatial and frequency (redshift) dependence of the 21-cm signal will reveal a full three dimensional picture of the IGM. Ultimately, the detection of this 21-cm signal will thus enable us to infer the ultra-violet (UV) and X-ray properties of the astrophysical sources responsible for reionisation.

Unfortunately, the cosmic 21-cm signal is extremely faint, buried beneath astrophysical foregrounds which can be up to five orders of magnitude brighter. However, this has not deterred numerous observational efforts to measure this elusive signal. Broadly speaking, these can be separated into two classes of experiments: (i) large-scale interferometer experiments designed to measure the spatial fluctuations in the cosmic 21-cm signal and (ii) global signal experiments which average the signal over the entire visible sky.

Several large-scale radio interferometers have been constructed or proposed. The first generation of these experiments, the Low-Frequency Array (LOFAR; \citealt{vanHaarlem:2013p200}), the Murchison Wide Field Array (MWA; \citealt{Tingay:2013p2997}), the Precision Array for Probing the Epoch of Reionisation (PAPER; \citealt{Parsons:2010p3000}) and the upgraded Giant Metrewave Radio Telescope (uGMRT; \citealt{Gupta:2017}) have limited sensitivities, requiring several years of observing time to potentially make a low signal-to-noise statistical detection of the signal. Second generation experiments, with considerably higher expected sensitivities, such as the Hydrogen Epoch of Reionization Array (HERA; \citealt{DeBoer:2017p6740}) and the Square Kilometre Array (SKA; \citealt{Mellema:2013p2975,Koopmans:2015}) should be able to provide higher signal-to-noise statistical detections across multiple redshifts. Further, the SKA has been designed to be able to provide the first three-dimensional tomographic images of the EoR.

Somewhat easier to design and operate, there have also been numerous global experiments. These include, the Experiment to Detect the Global EoR Signature (EDGES; \citealt{Bowman:2010p6724}), the Sonda Cosmol\'{o}gica de las Islas para la Detecci\'{o}n de Hidr\'{o}geno Neutro (SCI-HI; \citealt{Voytek:2014p6741}), the Shaped Antenna measurement of the background RAdio Spectrum (SARAS; \citealt{Patra:2015p6814}), Broadband Instrument for Global HydrOgen ReioNisation Signal (BIGHORNS; \citealt{Sokolowski:2015p6827}), the Large Aperture Experiment to detect the Dark Ages (LEDA; \citealt{Greenhill:2012p6829,Bernardi:2016p6834}), Probing Radio Intensity at high-Z from Marion (PRI$^{\rm Z}$M; \citealt{Philip:2019}) and the Netherlands-China Low-Frequency Explorer (NCLE\footnote{https://www.isispace.nl/projects/ncle-the-netherlands-china-low-frequency-explorer/})

Despite the wealth of experimental efforts to measure the cosmic 21-cm signal from the early Universe, there has only been one claimed detection during the cosmic dawn, an absorption feature in the global signal near $z\approx17$ announced by EDGES \citep{Bowman:2018}. However, the cosmological origin of this signal is heavily disputed in the literature \citep[see e.g.][]{Hills:2018,Draine:2018,Bowman:2018b,Bradley:2019,Singh:2019} and requires confirmation by an independent experiment.

All other attempts to measure the cosmic 21-cm signal have resulted in upper limits on the signal amplitude. For global experiments, these correspond to limits on the overall sky-averaged brightness temperature, with best available limits available from LEDA \citep{Bernardi:2016}, EDGES high-band \citep{Monsalve:2017} and SARAS2 \citep{Singh:2017}. In this work, we are specifically interested in the upper limits on the 21-cm power spectrum (PS), as this is a more information rich statistic.
 
The first upper limits on the 21-cm PS were measured by the Giant Metrewave Radio Telescope (GMRT; \citealt{Paciga:2013}) at $z\approx8.6$. Since then, each of the first generation of radio interferometers have all reported upper limits on the 21-cm PS. For LOFAR, these include single-night observations at $z=9.6-10.6$ \citep{Patil:2017} and more recently, upper limits at $z=19.8-25.2$ targeting the dark ages and cosmic dawn with the LOFAR-Low Band Antenna array \citep{Gehlot:2019}. Recently, the completed PAPER experiment revised their best upper limits \citep{Cheng:2018,Kolopanis:2019} across $z\approx7.5-11$ following identification of issues in dealing with signal loss from earlier reported limits \citep{Ali:2015p4327}. Finally, the MWA recently reported their best upper limits at $z=6.5-8.7$ from four seasons of observations \citep{Trott:2020}, improving on previous measurements in the literature \citep{Dillon:2015,Beardsley:2016,Barry:2019,Li:2019}. In addition to these, the Owens Valley Radio Observatory Long Wavelength Array (OVRO-LWA; \citealt{Eastwood:2019}) have also published upper limits at $z\approx18.4$.

With \citet{Mertens:2020}, LOFAR has now considerably improved their best upper limits on the EoR using 141 hours of observations at $z\approx9.1$ to achieve a $2\sigma$ upper limit on the 21-cm PS of $\Delta^{2}_{21}\approx(73~{\rm mK})^{2}$ at $k\approx0.075~h$~Mpc$^{-1}$. This corresponds to an improvement by a factor of $\approx8$ compared to their previous best upper limit \citep{Patil:2017}. While such limits are still several orders of magnitude above fiducial theoretical models \cite[e.g.][]{Mesinger:2016p6167}, they are aggressive enough to begin to rule out extreme models of reionisation known as `cold' reionisation \citep[e.g.][]{Mesinger:2013p1835,Parsons:2014p781}. Here, `cold' reionisation refers to models in which the IGM undergoes little to no heating, adiabatically cooling faster than the cosmic microwave background (CMB) as the Universe expands. This can result in large contrasts between the ionised and neutral IGM, driving 21-cm PS amplitudes in excess of $\Delta^{2}_{21}\gtrsim10^4~{\rm mK}^{2}$. Attempts to rule out such extreme regions of astrophysical parameter space have already been explored for EDGES High-band \citep{Monsalve:2017,Monsalve:2018,Monsalve:2019} and the now retracted upper limits from PAPER \citep{Pober:2015p4328,Greig:2016p6174}.

A related study using these new LOFAR upper limits was already performed by \citet{Ghara:2020}\footnote{While this work was nearing competition, a similar study by \citep{Mondal:2020} was published, focusing on placing limits on the contribution to the excess radio background from the high-$z$ Universe. These authors also used 21-cm PS emulators to explore a large parameter space of models. Under circumstances when the model assumptions are similar, our results are broadly consistent.}, using a different astrophysical model and semi-numerical simulation, \textsc{\small Grizzly} \citep{Ghara:2015,Ghara:2018}. Their focus was on general IGM properties, such as the mean IGM neutral fraction and spin temperature. For their Monte-Carlo Markov Chain (MCMC) results they trained a 21-cm PS emulator and emulators connecting the source properties to the IGM properties. In this work, we use a galaxy model \citep[e.g.][]{Park:2019} which allows us to directly compare against observations of the ultra-violet (UV) luminosity functions (LFs), as well as other observations of the first billion years. Thus, we place the recent LOFAR upper limits into context with other observational probes, finding that all models currently excluded by these new 21-cm upper limits are also excluded by existing probes, thereby strengthening each others individual evidence and solidly excluding extremely cold IGM models. Our framework directly forward models the 3D 21-cm signal.  Thus the inferred, marginalised IGM properties can serve as tests of emulator-based MCMCs, as well as confirming the robustness of the conclusions to the choice of simulation tool.

The outline of the remainder of this paper is as follows. In Section~\ref{sec:Method}, we summarise the astrophysical model used in this analysis, as well as the \cmmc{} setup. In Section~\ref{sec:results}, we discuss our main results and in Section~\ref{sec:Conclusion}, we provide our conclusions. Unless otherwise noted, all quoted quantities are in co-moving units with the following adopted cosmological parameters:  ($\Omega_\Lambda$, $\Omega_{\rm M}$, $\Omega_b$, $n$, $\sigma_8$, $H_0$) = (0.69, 0.31, 0.048, 0.97, 0.81, 68 km s$^{-1}$ Mpc$^{-1}$), consistent with recent results from the Planck mission \citep{PlanckCollaboration:2016p7780}.

\section{Methodology} \label{sec:Method}

\subsection{Simulating the 21-cm signal}

We simulate the cosmic 21-cm signal using the semi-numerical simulation code \cmfst{}\footnote{https://github.com/andreimesinger/21cmFAST}\citep{Mesinger:2007p122,Mesinger:2011p1123}. Specifically, we employ the most up-to-date astrophysical parameterisation \citep{Park:2019}, which allows the star-formation rate and ionising escape fraction to be dependent on the host dark matter halo. This, through simple conversions, allows \cmfst{} to directly compute UV LFs which can be compared against observed high-$z$ galaxy LFs. 

It is important to note that this parameterisation is a simplification of the true underlying astrophysics describing the ionising sources. It assumes only a single population of ionising sources (i.e. a single power-law relation with halo mass) and ignores any explicit redshift dependence on the escape fraction or stellar mass. Therefore, any conclusions drawn from this work are specific to the assumptions used in this source modelling. However, for the bulk of the galaxy population (i.e. $M_{UV} > -20$) the simple \citet{Park:2019} parameterisation is consistent with both observations of high-$z$ LFs as well as semi-analytical galaxy formation models and hydrodynamical simulations (e.g.\ \citealt{Harikane:2016,Mutch:2016,Xu:2016,Tacchella:2018,Behroozi:2019,Yung:2019}; Gillet et al., in prep). As such, there is currently no evidence that a more complicated source model is required at this stage. If more complex modelling is required, these existing parameters can be treated as effective, population averaged quantities. In the future, we can increase the source model complexity and use Bayesian evidence from our forward-modelling approach to quantify whether redshift-dependence is required by the observational data.

Finally, we adopt the on-the-fly ionising photon non-conservation correction (Park et al., in prep) which accounts for the fact that the excursion-set formalism used for tracking ionisations is not photon conserving in three dimensions \citep[e.g.][]{McQuinn:2005,Zahn:2007,Paranjape:2014,Paranjape:2016,Hassan:2017,Choudhury:2018,Hutter:2018,Molaro:2019}. Below we summarise the main ingredients of \cmfst{} and the corresponding astrophysical parameterisation, and refer the reader to the aforementioned works for further details.

\subsubsection{Galaxy UV properties}
It is assumed that the typical stellar mass of a galaxy, $M_{\ast}$, can be directly related to its host halo mass, $M_{\rm h}$ \citep[e.g.][]{Kuhlen:2012p1506,Dayal:2014b,Behroozi:2015p1,Mitra:2015,Mutch:2016,Sun:2016p8225,Yue:2016} through a power-law relation\footnote{At $z\gtrsim5$ this power-law dependence between stellar mass and halo mass is consistent with the mean relation from semi-analytic model predictions \citep[e.g][]{Mutch:2016,Yung:2019} and semi-empirical fits to observations \citep[e.g.][]{Harikane:2016,Tacchella:2018,Behroozi:2019}.} normalised to a dark matter halo of mass $10^{10}$~$M_{\odot}$:
\begin{eqnarray} \label{}
M_{\ast}(M_{\rm h}) = f_{\ast}\left(\frac{\Omega_{\rm b}}{\Omega_{\rm M}}\right)M_{\rm h},
\end{eqnarray}
where $f_{\ast}$ is the fraction of galactic gas in stars,
\begin{eqnarray} \label{}
f_{\ast} = f_{\ast, 10}\left(\frac{M_{\rm h}}{10^{10}\,M_{\odot}}\right)^{\alpha_{\ast}},
\end{eqnarray}
with $f_{\ast, 10}$ being the normalisation of the relation and $\alpha_{\ast}$ its power-law index.

The star-formation rate (SFR) can then be estimated by dividing the stellar mass by a characteristic time-scale,
\begin{eqnarray} \label{eq:Mdt}
\dot{M}_{\ast}(M_{\rm h},z) = \frac{M_{\ast}}{t_{\ast}H^{-1}(z)},
\end{eqnarray}
where $H^{-1}(z)$ is the Hubble time and $t_{\ast}$ is a free parameter allowed to vary between zero and unity.

Equivalently, the UV ionising escape fraction, $f_{\rm esc}$, is also parameterised to vary with halo mass,
\begin{eqnarray} \label{}
f_{\rm esc} = f_{\rm esc, 10}\left(\frac{M_{\rm h}}{10^{10}\,M_{\odot}}\right)^{\alpha_{\rm esc}},
\end{eqnarray}
with $f_{\rm esc, 10}$ again normalised at a halo of mass $10^{10}$~$M_{\odot}$ with the power-law index, $\alpha_{\rm esc}$.

Finally, to mimic the inability of small mass halos to host active, star-forming galaxies (i.e. because of inefficient cooling and/or feedback processes), a duty-cycle is included to suppress their contribution:
\begin{eqnarray} \label{}
f_{\rm duty} = {\rm exp}\left(-\frac{M_{\rm turn}}{M_{\rm h}}\right).
\end{eqnarray}
In effect, some fraction, $(1 - f_{\rm duty})$, of dark matter haloes with mass $M_{\rm h}$ are unable to host star-forming galaxies, with the characteristic scale for this suppression being set by $M_{\rm turn}$ \citep[e.g.][]{Shapiro:1994,Giroux:1994,Hui:1997,Barkana:2001p1634,Springel:2003p2176,Mesinger:2008,Okamoto:2008p2183,Sobacchi:2013p2189,Sobacchi:2013p2190}.

In summary, this results in six free parameters describing the UV galaxy properties in our model, $f_{\ast, 10}$, $f_{\rm esc, 10}$, $\alpha_{\ast}$, $\alpha_{\rm esc}$, $M_{\rm turn}$ and $t_{\ast}$.

\subsubsection{Galaxy X-ray properties}

Prior to reionisation, it is thought that the IGM undergoes heating in the early Universe due to X-rays. The likely, dominant source of these X-rays are stellar remnants within the first galaxies. In \cmfst{}, X-ray heating is included by calculating the cell-by-cell angle-averaged specific X-ray intensity, $J(\boldsymbol{x}, E, z)$, (in erg s$^{-1}$ keV$^{-1}$ cm$^{-2}$ sr$^{-1}$), by integrating the co-moving X-ray specific emissivity, $\epsilon_{\rm X}(\boldsymbol{x}, E_e, z')$ back along the light-cone:
\begin{equation} \label{eq:Jave}
J(\boldsymbol{x}, E, z) = \frac{(1+z)^3}{4\pi} \int_{z}^{\infty} dz' \frac{c dt}{dz'} \epsilon_{\rm X}  e^{-\tau},
\end{equation}
where $e^{-\tau}$ accounts for attenuation by the IGM. The co-moving specific emissivity, evaluated in the emitted frame, $E_{\rm e} = E(1 + z')/(1 + z)$, is, 
\begin{equation} \label{eq:emissivity}
\epsilon_{\rm X}(\boldsymbol{x}, E_{\rm e}, z') = \frac{L_{\rm X}}{\rm SFR} \left[ (1+\bar{\delta}_{\rm nl}) \int^{\infty}_{0}{\rm d}M_{\rm h} \frac{{\rm d}n}{{\rm d}M_{\rm h}}f_{\rm duty} \dot{M}_{\ast}\right],
\end{equation}
where $\bar{\delta}_{\rm nl}$ is the mean, non-linear density in a shell around $(\boldsymbol{x}, z)$ and the quantity in square brackets is the SFR density along the light-cone. 

The normalisation, $L_{\rm X}/{\rm SFR}$ (erg s$^{-1}$ keV$^{-1}$ $M^{-1}_{\odot}$ yr), is the specific X-ray luminosity per unit star formation escaping the host galaxies. This is assumed to be a power-law with respect to photon energy, $L_{\rm X} \propto E^{- \alpha_{\rm X}}$, which is attenuated below a threshold energy, $E_0$, where photons are absorbed inside the host galaxy. The specific luminosity is then normalised to an integrated soft-band ($<2$~keV) luminosity per SFR (in erg s$^{-1}$ $M^{-1}_{\odot}$ yr), which is taken to be a free parameter:
\begin{equation} \label{eq:normL}
  L_{{\rm X}<2\,{\rm keV}}/{\rm SFR} = \int^{2\,{\rm keV}}_{E_{0}} dE_e ~ L_{\rm X}/{\rm SFR} ~.
\end{equation}
This limit of $2\,{\rm keV}$ approximates to an X-ray mean-free path of roughly the Hubble length at high redshifts, implying harder photons do not contribute to heating the IGM \citep[e.g.][]{McQuinn:2012p3773}.

In summary, we have three free parameters describing the X-ray properties,  $L_{{\rm X}<2\,{\rm keV}}/{\rm SFR}$ , $E_{0}$ and $\alpha_{\rm X}$.

\subsubsection{Ionisation and thermal state of the IGM}

The evolved density and velocity fields in \cmfst{} are obtained following second-order Lagrange perturbation theory \citep[e.g][]{Scoccimarro:1998p7939}. Reionisation is then determined from the evolved density field by tracking the balance between the cumulative number of ionising photons against the number of neutral hydrogen atoms plus cumulative recombinations in spheres of decreasing radii. Each cell is flagged as ionised when,
\begin{eqnarray} \label{eq:ioncrit}
n_{\rm ion}(\boldsymbol{x}, z | R, \delta_{R}) \geq (1 + \bar{n}_{\rm rec})(1-\bar{x}_{e}),
\end{eqnarray}
where $\bar{n}_{\rm rec}$ is the cumulative number of recombinations \citep[e.g.][]{Sobacchi:2014p1157} and $n_{\rm ion}$ is the cumulative number of IGM ionising photons per baryon inside a spherical region of size, $R$ and corresponding overdensity, $\delta_{R}$,
\begin{eqnarray} \label{eq:ioncrit2}
n_{\rm ion} = \bar{\rho}^{-1}_b\int^{\infty}_{0}{\rm d}M_{\rm h} \frac{{\rm d}n(M_{h}, z | R, \delta_{R})}{{\rm d}M_{\rm h}}f_{\rm duty} \dot{M}_{\ast}f_{\rm esc}N_{\gamma/b},
\end{eqnarray}
where $\bar{\rho}_b$ is the mean baryon density and $N_{\gamma/b}$ is the number of ionising photons per stellar baryon\footnote{We adopt a value of 5000, corresponding to a Salpeter initial mass function \citep{Salpeter:1955}; however note that this is highly degenerate with $f_{\ast}$}.
The final term of Equation~\ref{eq:ioncrit}, $(1-\bar{x}_{e})$, corresponds to ionisations by X-rays, which are expected to contribute at the $\sim 10$ per cent level \citep[e.g.][]{Ricotti:2004p7145,Mesinger:2013p1835,Madau:2017,Ross:2017,Eide:2018}.

The thermal state of the neutral IGM (and its partial ionisations) are tracked in each cell by following adiabatic heating/cooling, Compton heating/cooling, heating through partial ionisations, as well as the heating/ionisations from X-rays (discussed in the previous section). Combining all these, the IGM spin temperature, $T_{\rm S}$, is then computed as a weighted mean between the gas and CMB temperatures, depending on the density and local Lyman-$\alpha$ (Ly$\alpha$) intensity impinging on each cell \citep{Wouthuysen:1952p4321,Field:1958p1}.

This Ly$\alpha$ background (see e.g. \citealt{Mesinger:2011p1123} for further details) is estimated by summing the contribution from: (i) excitations of neutral hydrogen by X-rays ($J_{\alpha,{\rm X}}$) and (ii) direct stellar emission of photons between Ly$\alpha$ and the Lyman limit ($J_{\alpha,\ast}$). 
For (i), this is set by the X-ray heating rate assuming that energy injection is balanced by photons redshifting out of Ly$\alpha$ resonance \citep{Pritchard:2007}. For (ii), any Lyman-$n$ resonance photons absorbed by the neutral IGM cascade with a recycling fraction passing through Ly$\alpha$ producing a background that is the sum over all Lyman resonances \citep[e.g.][]{Barkana:2005p1934}. Note, the soft UV spectra of the stellar emission component of the first sources is currently held fixed and we do not include other possible sources of soft UV, such as quasars \citep[see e.g.][]{Qin:2017,Ricci:2017,Mitra:2018}.

\subsubsection{Ionising photon non-conservation correction}

As \cmfst{} employes an excursion-set formalism for determining which cells are ionised, it is impacted by the resultant non-conservation of ionising photons. This arises within cells that exceed the ionisation criteria (Eq.~\ref{eq:ioncrit}), where the excess ionising photons that are left over after ionising the neutral hydrogen are not propagated onward. This behaviour acts as an effective bias on the ionising escape fraction, $f_{\rm esc}$. Roughly speaking, this corresponds to a loss of $\sim10-20$ per cent of the ionising photons \citep[e.g.][]{McQuinn:2005,Zahn:2007,Paranjape:2014,Paranjape:2016,Hassan:2017,Choudhury:2018,Hutter:2018,Molaro:2019}. 

Both \citet{Choudhury:2018} and \citet{Molaro:2019} have introduced new, explicitly photon conserving algorithms for semi-numerical simulations. While being orders of magnitude faster than full radiative-transfer simulations, they are still considerably slower than conventional semi-numerical simulations. Under Bayesian parameter estimation approaches such as \cmmc{}, these schemes are still intractable when forward modelling the 21-cm signal in the high-dimensional parameter spaces required to characterise the ionising, soft UV, and X-ray properties of the first galaxies.

Instead, Park et al., in prep, introduce an approximate correction to the ionising photon non-conservation issue, correcting for the effective bias on $f_{\rm esc}$ by analytically solving for the correct evolution of the ionisation fraction for a given source model, under the assumption of no correlations between the sources and sinks (which should only impact the final $\sim10$ per cent of the EoR; \citealt{Sobacchi:2014p1157}).

This analytic solution is compared against a calibration curve from \cmfst{} considering only ionisations (i.e. no recombinations or spin temperature evolution). If photons were conserved, these two curves would be identical. However, in practice the \cmfst{} history is delayed due to the loss of ionising photons. In order to recalibrate \cmfst{}, we decrease the redshift used for determining the ionisation field\footnote{This effectively amounts to boosting the number of ionisations to compensate for the ionising photon non-conservation. However, we do not know $a$-priori how much to boost ionisations by, thus we modify the redshift to allow our calculation to be performed on-the-fly.}. Note, this modified redshift is only used for modifying the ionisation field, all other quantities are calculated at the original redshift. We find that for the bulk of the EoR this corresponds roughly to a shift in redshift of $\Delta z\sim0.3 \pm 0.1$.

\subsubsection{Computing the 21-cm signal}

Combining all the cosmological fields discussed in the previous section, we compute the observable cosmic 21-cm signal as a brightness temperature contrast relative to the CMB temperature,  $T_{\rm CMB}$ \citep[e.g.][]{Furlanetto:2006p209}:
\begin{eqnarray} \label{eq:21cmTb}
\delta T_{\rm b}(\nu) &=& \frac{T_{\rm S} - T_{\rm CMB}(z)}{1+z}\left(1 - {\rm e}^{-\tau_{\nu_{0}}}\right)~{\rm mK},
\end{eqnarray}
where $\tau_{\nu_{0}}$ is the optical depth of the 21-cm line,
\begin{eqnarray}
\tau_{\nu_{0}} &\propto& (1+\delta_{\rm nl})(1+z)^{3/2}\frac{x_{\hi{}}}{T_{\rm S}}\left(\frac{H}{{\rm d}v_{\rm r}/{\rm d}r+H}\right).
\end{eqnarray}
Here, $x_{\hi{}}$ is the neutral hydrogen fraction, $\delta_{\rm nl} \equiv \rho/\bar{\rho} - 1$ is the gas over-density, $H(z)$ is the Hubble parameter, ${\rm d}v_{\rm r}/{\rm d}r$ is the gradient of the line-of-sight component of the velocity and $T_{\rm S}$ is the gas spin temperature. All quantities are evaluated at redshift $z = \nu_{0}/\nu - 1$, where $\nu_{0}$ is the 21-cm frequency and we drop the spatial dependence for brevity. Additionally, we include the impact of redshift space distortions along the line-of-sight as outlined in \citet{Mao:2012p7838,Jensen:2013p1389,Greig:2018}.

\subsection{Astrophysical parameter set} \label{sec:fiducial}

Under the assumptions of the adopted astrophysical model described previously, we have nine astrophysical parameters. Below, we summarise each of these and the associated parameter ranges we adopt based on previous explorations with \cmmc{} \citep{Greig:2017,Park:2019}. We additionally summarise these in the top row of Table~\ref{tab:Results}. Within \cmmc{} we adopt flat priors over all of these parameter ranges.
\begin{itemize}
\item[(i)]$f_{\ast, 10}$: normalisation for the fraction of galactic gas in stars evaluated at a halo mass of 10$^{10}~M_{\odot}$. This is allowed to vary in the log as, ${\rm log}_{10}(f_{\ast, 10}) \in [-3,0]$.
\item[(ii)]$\alpha_{\ast}$: power-law index for the star-formation as a function of halo mass, allowed to vary between, $\alpha_{\ast} \in [-0.5,1]$.
\item[(iii)]$f_{\rm esc, 10}$: normalisation for the ionising UV escape fraction evaluated at a halo mass of 10$^{10}~M_{\odot}$. We vary this in the log range, ${\rm log}_{10}(f_{\rm esc, 10}) \in [-3, 0]$.
\item[(iv)]$\alpha_{\rm esc}$: power-law index for the ionising UV escape fraction as a function of halo mass, allowed to vary in the range, $\alpha_{\rm esc} \in [-1,0.5]$.
\item[(v)]$t_{\ast}$: the star-formation time scale as a fraction of the Hubble time, allowed to vary in the range, $t_{\ast} \in (0,1]$.
\item[(vi)]$M_{\rm turn}$: halo mass turn-over below which the abundance of active star-forming galaxies is exponentially suppressed by the adopted duty cycle. This is allowed to vary within the range, ${\rm log}_{10}(M_{\rm turn}) \in[8,10]$.
\item[(vii)]$E_{0}$: the minimum energy threshold for X-ray photons capable of escaping their host galaxy. Allowed to vary within the range, $E_{0} \in [0.2,1.5]$~keV, which corresponds of an integrated column density of, ${\rm log_{10}}(N_{\hi{}}/{\rm cm}^{2}) \in [19.3,23.0]$.
\item[(viii)]$L_{{\rm X}<2\,{\rm keV}}/{\rm SFR}$: the normalisation for the soft-band X-ray luminosity per unit star-formation determined over the $E_{0} - 2$~keV energy band. Allowed to vary across, ${\rm log_{10}}(L_{{\rm X}<2\,{\rm keV}}/{\rm SFR}) \in [30, 42]$. Here, we considerably reduce the lower bound for $L_{{\rm X}<2\,{\rm keV}}/{\rm SFR}$ compared to what is typically chosen \citep[i.e.][]{Park:2019} in-order to explore truly `cold' reionisation scenarios.
\item[(iv)]$\alpha_{\rm X}$: the power-law index for the SED of X-ray sources, which we allow to vary between $\alpha_{\rm X} \in [-1,3]$.
\end{itemize}

\subsection{\cmmc{} setup} \label{sec:setup}

\cmmc{}\footnote{https://github.com/BradGreig/21CMMC} is a massively parallel MCMC sampler of 3D semi-numerical reionisation simulations \citep{Greig:2015p3675,Greig:2017p8496,Greig:2018,Park:2019}. It is based on the \textsc{\small Python} module \textsc{\small CosmoHammer} \citep{Akeret:2012p842} which uses the \textsc{\small Emcee} \textsc{\small Python} module \citep{ForemanMackey:2013p823}, an affine invariant ensemble sampler from \citet{Goodman:2010p843}. At each proposal step, \cmmc{} performs an independent 3D realisation of the 21-cm signal using \cmfst{} to obtain a sampled 21-cm PS. 

Typically, we would perform our MCMC seeking to maximise the likelihood that our astrophysical model is consistent with the observational data. However, current upper limits on the 21-cm signal are not strong enough to yield statistically significant constraints on the astrophysical parameter space. Instead, following \citet{Ghara:2020}, we invert the typical problem by reversing the likelihood, maximising models which exceed the observational limits\footnote{This approach corresponds to identifying models ruled out by LOFAR in order to allow our forward modelling approach to converge quicker.}. This enables us to explore the regions of astrophysical parameter space inconsistent with the recent LOFAR upper limits at $z\approx9.1$. In this work, we model the likelihood as a one-sided Gaussian\footnote{Our choice is motivated by the fact that we treat the measured data-points quoted in \citet{Mertens:2020} as limits. This differs from \citet{Ghara:2020} who adopt an error function centred on the actual data-point value. As a result, our likelihood is more conservative than that of \citet{Ghara:2020}, however, ultimately the differences between these approaches is negligible.} for each of the Fourier, $k$, data points quoted in \citet{Mertens:2020}. Therefore, our likelihood\footnote{This choice of zero is motivated by the fact that we model our likelihood as a one-sided Gaussian. Any model below the upper limit is perfectly consistent with the data, and thus would normally have a likelihood of unity. However, since we are interested in models which instead exceed the upper limits, we then subtract by unity resulting in zero likelihood. Note, in practice we only evaluate the likelihood at $k\approx0.075~h~{\rm Mpc}^{-1}$ rather than evaluating the full product. This is motivated by the fact that all \cmfst{} models were found to be below the upper-limit at the second $k$-bin.}, $\mathcal{L}$, is a product over all Fourier bins, $k$:
\begin{eqnarray}
\mathcal{L} = \prod^{n}_{i} \left\{ \begin{array}{cl} 
0  &  \Delta^{2}_{21, m} < \Delta^{2}_{21,d} \\
1 - {\rm exp}\left[ -\frac{1}{2} \left( \frac{ \Delta^{2}_{21,m} - \Delta^{2}_{21,d}}{\sigma_{i}}\right)^{2}\right] & \Delta^{2}_{21, m} \geq \Delta^{2}_{21,d},
\end{array} \right.
\end{eqnarray}
where $\Delta^{2}_{21, m}$ and $\Delta^{2}_{21,d}$ correspond to the 21-cm PS from the astrophysical model sampled in \cmmc{} and the LOFAR upper limits, respectively\footnote{Note that this assumes the sample variance error is Gaussian distributed, which is a reasonable approximation in the modest S/N regime relevant for current and upcoming observations \citep{Mondal:2017,Shaw:2020}.}. Note, we have dropped the Fourier dependence from these terms, i.e. $\Delta^{2}_{21, m} = \Delta^{2}_{21, m}(k_{i})$, for brevity.
 This likelihood is then interpreted as:
\begin{itemize}
\item astrophysical models below the upper limit (i.e. $\Delta^{2}_{21, m} < \Delta^{2}_{21,d}$) are perfectly consistent with the LOFAR upper limits and are thus given a likelihood of zero.
\item above the upper limit (i.e. $\Delta^{2}_{21, m} \geq \Delta^{2}_{21,d}$) the likelihood smoothly transitions from zero following a one-sided Gaussian of width equal to the total uncertainty, $\sigma$, on the upper limit measurement. This implies astrophysical models with amplitudes much larger than the limit are ruled-out with greater certainty than those close to the central-limit.
\end{itemize}

The total uncertainty, $\sigma$ on the upper limit is the quadrature sum of the quoted uncertainty on the limit value from \citet{Mertens:2020} and a conservative 20 per cent multiplicative modelling uncertainty on the sampled 21-cm PS within \cmmc{}. This modelling uncertainty is motivated by approximations adopted in semi-numerical simulations relative to radiative-transfer simulations \citep[e.g.][]{Zahn:2011p1171,Ghara:2018,Hutter:2018}.

The largest scale probed by the LOFAR upper limits corresponds to $k\approx0.075~h~{\rm Mpc}^{-1}$, which is a factor of two larger than the typical sampling scale adopted in \cmmc{} (i.e. $k=0.1-1.0$~Mpc$^{-1}$). Therefore, to adequately sample these much larger spatial scales, within \cmmc{} we perform 3D realisations of the cosmic 21-cm signal in a box of 400~Mpc and 128 voxels per side-length\footnote{For our simulation volume, we estimate a sample variance of $\sim13$ per cent for the largest physical scale sampled by LOFAR (i.e. $k\approx0.075~h~{\rm Mpc}^{-1}$), which is well below that of the combined error from the observations and modelling uncertainty.}. This corresponds to a resolution that is $\sim3$ Mpc per voxel, slightly larger than the preferred resolution of \cmfst{}. However, we have performed convergence tests to verify that the increased resolution of $\sim3$ Mpc per voxel does not alter the results.

\section{Results} \label{sec:results}

\subsection{Disfavoured parameters} \label{sec:astro}

\begin{figure*} 
	\begin{center}
	  \includegraphics[trim = 0.6cm 0.8cm 0cm 0cm, scale = 0.5]{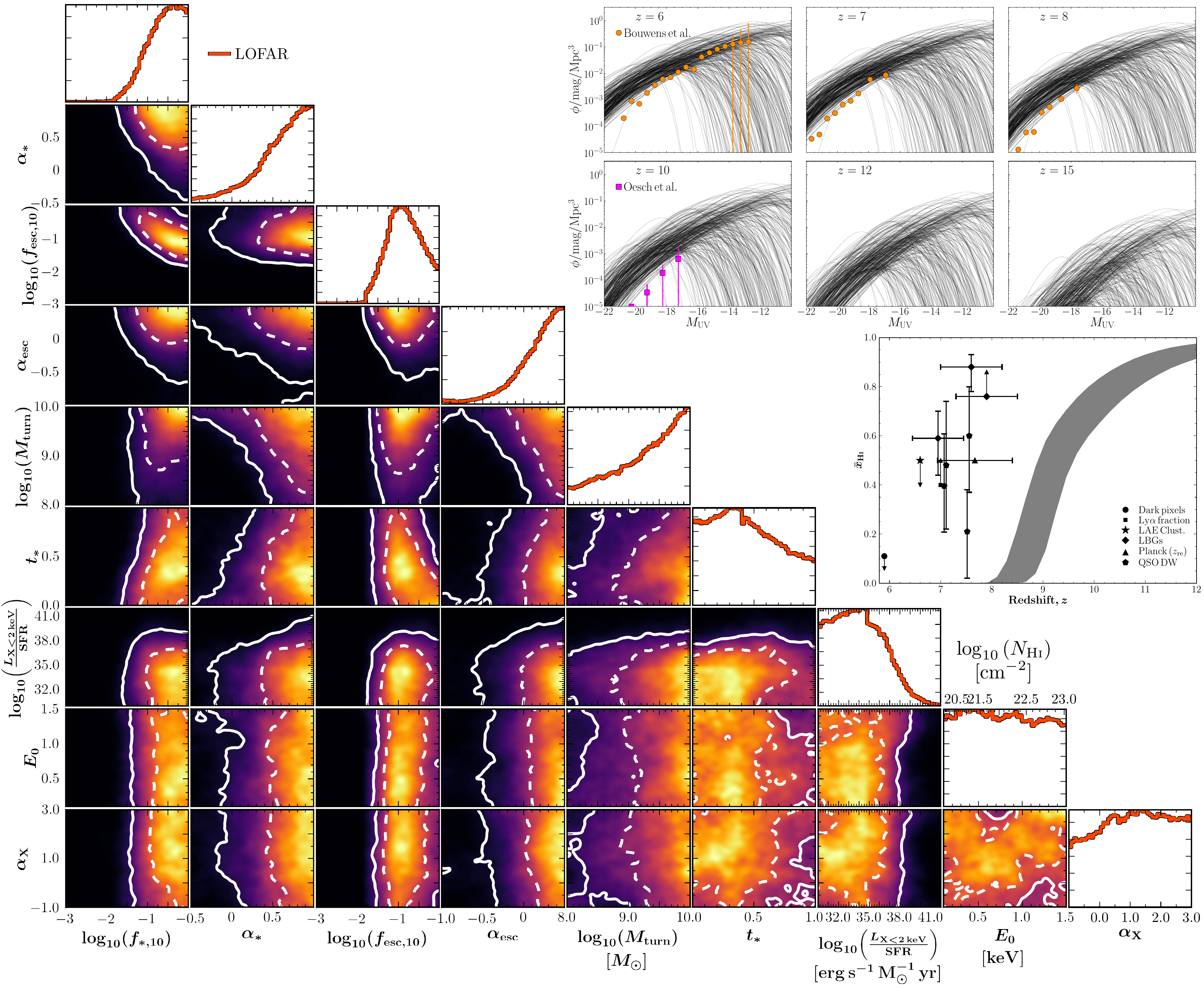}
	\end{center}
\caption[]{Marginalised one and two dimensional posterior distributions for the parameters which are disfavoured by just the LOFAR upper limits (i.e. no other observational probes) at $z\approx9.1$ \citep{Mertens:2020}. White dashed (solid) contours correspond to the 68th (95th) percentiles. In the top right panels, we provide 500 randomly sampled LFs which are drawn from the posterior of models that exceed the LOFAR upper limits and are compared against existing constraints at $z=6-8$ \citep{Bouwens:2015,Bouwens:2017} and $z=10$ \citep{Oesch:2018}. In the middle right panel, we compare the bounds on the reionisation histories disfavoured by LOFAR to all current observational constraints on the IGM neutral fraction (see text for further details).}
\label{fig:Corner}
\end{figure*}

\begin{table*}
\tiny
\begin{tabular}{@{}lccccccccc}
\hline
  & ${\rm log_{10}}(f_{\ast,10})$ & $\alpha_{\ast}$ & ${\rm log_{10}}(f_{\rm esc,10})$ & $\alpha_{\rm esc}$ & ${\rm log_{10}}(M_{\rm turn})$ & $t_{\ast}$ & ${\rm log_{10}}\left(\frac{L_{{\rm X}<2{\rm keV}}}{\rm SFR}\right)$ & $E_0$  & $\alpha_{\rm X} $ \\
               &  &  &  &  & $[{\rm M_{\sun}}]$ & &  $[{\rm erg\,s^{-1}\,M_{\sun}^{-1}\,yr}]$ &  $[{\rm keV}]$ \\
\hline
\vspace{0.8mm}
Prior ranges & [-3.0, 0.0] & [-0.5,1.0] & [-3.0, 0.0] & [-1.0,0.5] & [8.0,10.0] & (0.0,1.0] & [30.0,42.0] & [0.2,1.5] & [-1.0,3.0] \\
\hline
\vspace{0.8mm}
68th percentile limits & [-1.06,0.0] &  [0.28,1.0]  & [-1.27, -0.39]  &  [-0.11, 0.5]  &  [8.66, 10.0]  &  [0.0, 0.76]  &   [30, 36.58]   & [0.2,1.5] & [-0.2, 3.0]  \\
95th percentile limits & [-1.53,0,0] &  [-0.23,1.0]  & [-1.64, 0.0]  &   [-0.64, 0.5] &  [8.13, 10.0] &    [0.0, 0.96]  &   [30, 39.04]   &  [0.2,1.5] & [-0.87, 3.0]\\
\hline
\end{tabular}
\caption{Summary of the 68th and 95th percentile limits on the disfavoured regions of nine astrophysical parameters included in \cmmc{} using the LOFAR upper limits at $z\approx9.1$ from \citet{Mertens:2020}.}
\label{tab:Results}
\end{table*} 

In Figure~\ref{fig:Corner}, we present the marginalised one and two dimensional posterior distributions output from \cmmc{} using the setup described in Section~\ref{sec:setup} for the nine input parameters outlined in Section~\ref{sec:fiducial}. Note that these results are obtained using only the LOFAR upper limits, no other existing observational constraints are used. The resultant 68th and 95th percentile limits on these parameters disfavoured by the LOFAR limits are summarised in Table~\ref{tab:Results}.

It is imperative to point out that here we are only exploring models which exceed the LOFAR upper limits at $z\approx9.1$. Therefore, any non-zero regions of the marginalised posteriors correspond to models which are in excess of the LOFAR upper limits. This approach is more of a demonstration that: (i) LOFAR is already capable of ruling out some models and (ii) that forward modelled MCMC tools during the EoR such as \cmmc{} are necessary for interpreting the results. Nevertheless, below we discuss some of the deductions that can be made from the existing LOFAR upper limits. Note again though, that these results are specific to the underlying assumptions of our source model.

The most interesting limit that we can discern from Figure~\ref{fig:Corner} is that of the soft-band X-ray luminosity, $L_{{\rm X}<2\,{\rm keV}}/{\rm SFR}$. This directly controls the amount of heating that the IGM undergoes between the dark ages and reionisation and is responsible for the `cold' reionisation scenarios. At 95 per cent confidence, we are able to place limits of ${\rm log_{10}}\left(L_{{\rm X}<2\,{\rm keV}}/{\rm SFR}\right) \lesssim 39.04$. These limits sit below those from local populations of star-forming galaxies \citep{Mineo:2012p6282}, stacked {\it Chandra} observations \citep{Lehmer:2016p7810} and predictions at high-redshifts by population synthesis models \citep{Fragos:2013p6529}. For the remaining two galaxy X-ray properties, $E_0$ and $\alpha_{\rm X}$ we are currently unable to place any meaningful limits. This is driven by the fact that we are ruling out models with little to no X-ray heating. In the absence of X-ray heating, it does not matter how much the X-ray SED is attenuated or its power-law shape.

For the galaxy UV properties, the limits are driven by the requirement to maximise the 21-cm PS amplitude at $z\approx9.1$ (to be able to exceed the LOFAR limit). This occurs when we are roughly at the mid-point of reionisation (i.e. $\bar{x}_{\hi{}} \sim 0.5$) where the 21-cm PS peak is typically\footnote{The peak amplitude of the 21-cm PS is actually strongly model dependent and need not occur at the mid-point of reionisation. However, throughout we refer to the peak occurring near the mid-point of reionisation purely for the purpose of making a generalised qualitative statement.} at its maximum \citep[e.g.][]{Mellema:2006,Lidz:2008p1744}. Therefore, the limits on $f_{\ast,10}$, $\alpha_{\ast}$, $f_{\rm esc,10}$, $\alpha_{\rm esc}$ and $M_{\rm turn}$ are completely degenerate. For $f_{\rm esc,10}$ we recover constraints on disallowed values, but for the remainder we only recover limits. This arises due to the completely degenerate nature between $f_{\ast,10}$ and $f_{\rm esc,10}$ for single redshift measurements and the absence of any other observational constraints (i.e. UV LFs, \citealt{Park:2019}). Further contributing to this are the different prior ranges for the corresponding power-law indices $\alpha_{\ast}$ and $\alpha_{\rm esc}$. Nevertheless, these limits imply that LOFAR disfavours normalisations of the mass dependent star-formation rate, $f_{\ast,10}$ and escape fraction, $f_{{\rm esc},10}$, to be above $\sim3.1$ and $\sim2.3$ per cent, respectively. Also, a minimum source mass above ${\rm log_{10}}(M_{\rm turn}) \gtrsim 8.13$ is disfavoured at 95 per cent confidence.

Unfortunately, owing to differences in the choice of varied parameters (and their allowed ranges), their corresponding meaning (i.e. the mass dependence of star-formation rates and escape fractions in \cmfst{}), direct comparison to the results of \citet{Ghara:2020} is not possible. However, the general trends of both this work and of \citet{Ghara:2020} are consistent: (i) EoR to be ongoing at $z\approx9.1$ disfavouring low mass star-forming galaxies and high ionising escape fractions and (ii) very low X-ray luminosities.

\subsection{Comparison against existing observations}

Next, we focus more on globally averaged quantities and how they compare against existing independent observations and limits in the literature.

\subsubsection{Reionisation history}

In the middle right panel of Figure~\ref{fig:Corner} we present a census of all existing constraints on the IGM neutral fraction against the full range of reionisation histories disfavoured by LOFAR (shaded region). We compare against the dark pixel statistics of high-$z$ quasars (QSOs; \citealt{McGreer:2015p3668}), Ly$\alpha$ fraction \citep{Mesinger:2015p1584}, the clustering of Ly$\alpha$ emitters (LAEs; \citealt{Sobacchi:2015}), the Ly$\alpha$ equivalent width distribution of Lyman-break galaxies (LBGs; \citealt{Mason:2018,Hoag:2019,Mason:2019}), the neutral IGM damping wing imprint from high-$z$ QSOs \citep{Greig:2017,Davies:2018,Greig:2018} and the midpoint of reionisation ($z_{\rm Re}$) from Planck \citep{Planck:2018}. 

As discussed in the previous section, the models disfavoured by the LOFAR upper limits are driven to produce a midpoint of reionisation at $z\approx9.1$. The full range of IGM neutral fractions disfavoured by LOFAR at $z\approx9.1$ correspond to $0.15 \lesssim \bar{x}_{\hi{}} \lesssim 0.6$. Compared to the existing observational constraints, this places these disfavoured models at roughly $\gtrsim2\sigma$. Unfortunately, owing to the still large amplitude of the LOFAR upper limits, and the correspondingly large uncertainties, we have verified that a joint analysis is entirely constrained by the existing limits. However, prospects for this will improve in the near future with the further processing of existing observational data and multiple frequency (redshift) limits.

Our constraints on $\bar{x}_{\hi{}}$ are broadly consistent with \citet{Ghara:2020}, with any differences being explained by different choices in adopted priors between the two works. The \citet{Ghara:2020} limits on $\bar{x}_{\hi{}}$ appear from two regions of parameter space (c.f their Table 5): (i) `cold' reionisation ($0.40 < \bar{x}_{\hi{}} < 0.55$) and (ii) patchy X-ray heating ($\bar{x}_{\hi{}}>0.92$). For (i), reionisation must be ongoing {\it and} the IGM must be cold. Our results are consistent with (i), given that \citet{Ghara:2020} assume a fixed $M_{\rm turn}=10^{9}~M_{\odot}$ while also including a hard prior on the neutral fraction excluding values below 0.19, while we have a flat prior down to zero. For (ii), \citet{Ghara:2020} require reionisation to be in its infancy {\it and} X-ray heating to be dominated by luminous, highly biased ($M_{h} \geq 10^{10}~M_{\odot}$) sources with very soft SEDs ($E_{0}=0.2$~keV). We do not recover their (ii) because our prior on $M_{\rm turn}$ has a hard upper limit of $M_{h} \leq 10^{10}~M_{\odot}$.

\subsubsection{UV luminosity functions}

Unlike the reionisation history which is constrained to be within a narrow range around $\bar{x}_{\hi{}}\approx0.5$, the UV LFs are not strongly constrained. Instead, they show extremely large scatter in the shapes and amplitudes, especially at the faint end. As such, we cannot simply provide a constrained range of plausible UV LFs as we did previously for the reionisation history (shaded region of middle right panel of Figure~\ref{fig:Corner}). Therefore, in the six panels in the top right of Figure~\ref{fig:Corner} we plot the resultant UV luminosity functions at $z=6$, 7, 8, 10, 12 and 15 for 500 randomly sampled models drawn from the posterior which exceed the LOFAR upper limit. This is one of the advantages of using \cmfst{} in that the built-in  parameterisation is capable of directly outputting UV LFs.

We compare these UV LFs against existing observations from unlensed UV LFs at $z=6-8$ (orange circles; \citealt{Bouwens:2015,Bouwens:2017}) and at $z=10$ (pink squares; \citep{Oesch:2018}). The vast majority of the UV LFs disfavoured by LOFAR are strongly disfavoured by the existing UV galaxy limits (up to two orders of magnitude larger in amplitude). Again, this is driven by models producing excessive amounts of ionising photons to have reionisation $\sim50$ per cent complete by $z\approx9.1$. While the UV LFs are independent of the number of ionising photons, to achieve cold reionisation at $z\approx9.1$ we prefer to have both a high $f_{\rm esc}$ and high stellar-to-halo mass relation (see the $f_{\ast}  - f_{\rm esc}$ panel in Figure~\ref{fig:Corner}). Interestingly, unlike the existing constraints on the IGM neutral fraction, there appear to be several models in excess of the LOFAR limits which appear to be capable of producing UV LFs consistent with the observed ones. However, the vast majority are strongly inconsistent, owing to the relatively small statistical uncertainties on the observed UV LFs.

Once again, this highlights the fundamental value of observing the cosmic 21-cm signal. Constraints on the amplitude of the 21-cm PS are capable of providing limits on the underlying UV galaxy LFs, beyond what is capable from existing space telescopes. Our galaxy model, which has an average star-formation time-scale evolving with redshift as the Hubble time (or analogously the halo dynamical time; Equation~\ref{eq:Mdt}), and a redshift-independent stellar-to-halo mass relation is capable of reproducing current observations of high-$z$ ($z$~6-10) UV LFs (e.g. figure 4 of \citealt{Park:2019}). Moreover, these scaling relations are consistent with more sophisticated high-$z$ SAMs \citep[e.g.][]{Mutch:2016,Yung:2019}, as well as empirical scaling relations \citep[e.g.][]{Tacchella:2018}. Therefore the fact that the UV LFs corresponding to the models inconsistent with the LOFAR observations are generally above the observed UV LFs is a significant result, highlighting that such models are already ruled out by current observations.

\subsubsection{Electron scattering optical depth, $\tau_{e}$}

\begin{figure} 
	\begin{center}
	  \includegraphics[trim = 1.2cm 0.8cm 0cm 0cm, scale = 0.48]{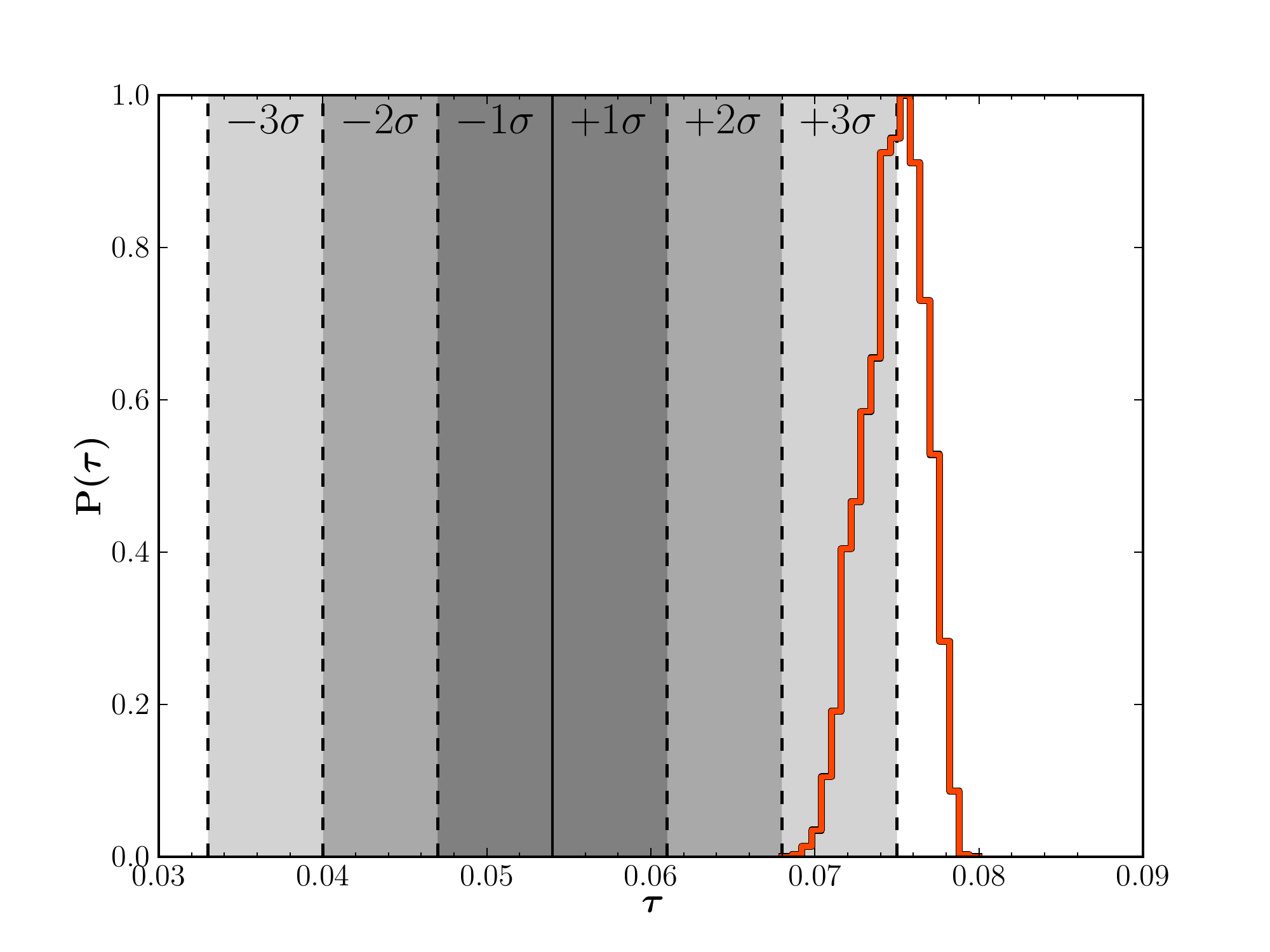}
	\end{center}
\caption[]{Histogram (red curve) of $\tau_{\rm e}$ from all models found to be in excess of the LOFAR upper limits on the 21-cm PS at $z\approx9.1$. Shaded bands correspond to the statistical uncertainty on $\tau_{\rm e}$ as measured by Planck ($\tau_{\rm e} = 0.054 \pm 0.007$; \citealt{Planck:2018}).}
\label{fig:Tau}
\end{figure}

Next, in Figure~\ref{fig:Tau} we consider the electron scattering optical depth, $\tau_{\rm e}$, using the latest constraints from Planck ($\tau_{\rm e} = 0.054 \pm 0.007$; \citealt{Planck:2018}). Here, the solid vertical line is the mean value from Planck, with shaded regions corresponding to being within $\pm1\sigma$, $\pm2\sigma$, $\pm3\sigma$ of the mean observed value. The red curve is the histogram of $\tau_{\rm e}$ obtained by binning all models disfavoured by the LOFAR upper limits. Clearly, all models in excess of LOFAR are $\gtrsim2\sigma$ from existing observational constraints. Note again, this arises owing to the assumptions of our astrophysical source parameterisation\footnote{We note that our parametrisation does allow for the {\it population averaged} $f_{\rm esc}$ to vary with redshift. This is because the halo mass function evolves with redshift, and we allow $f_{\rm esc}$ to scale with the halo mass through a power law relation whose index, $\alpha_{\rm esc}$, is a free parameter.  However, our range of priors on $\alpha_{\rm esc}$ results in a relatively modest redshift evolution of the population averaged galaxy $f_{\rm esc}$.}. For example, allowing a strong redshift dependence on $f_{\rm esc}$ could produce models consistent with Planck. However, we note that most hydrodynamical simulations do not find evidence for a strong redshift evolution for $f_{\rm esc}$ (e.g.\ \citealt{Kimm:2014,Paardekooper:2015,Xu:2016,Ma:2020}, though see \citealt{Lewis:2020}). The primary driver of this is the requirement to have reionisation at $\sim50$ per cent by $z\sim9.1$ for the 21-cm signal to exceed the current LOFAR limits, whereas the observed values from Planck prefer a redshift for the mid-point of $z_{\rm Re} = 7.67 \pm 0.5$ \citealt{Planck:2018}). 

\subsubsection{Mean UV photo-ionisation rate, $\bar{\Gamma}_{\rm UVB}$}

\begin{figure} 
	\begin{center}
	  \includegraphics[trim = 1.2cm 0.8cm 0cm 0cm, scale = 0.48]{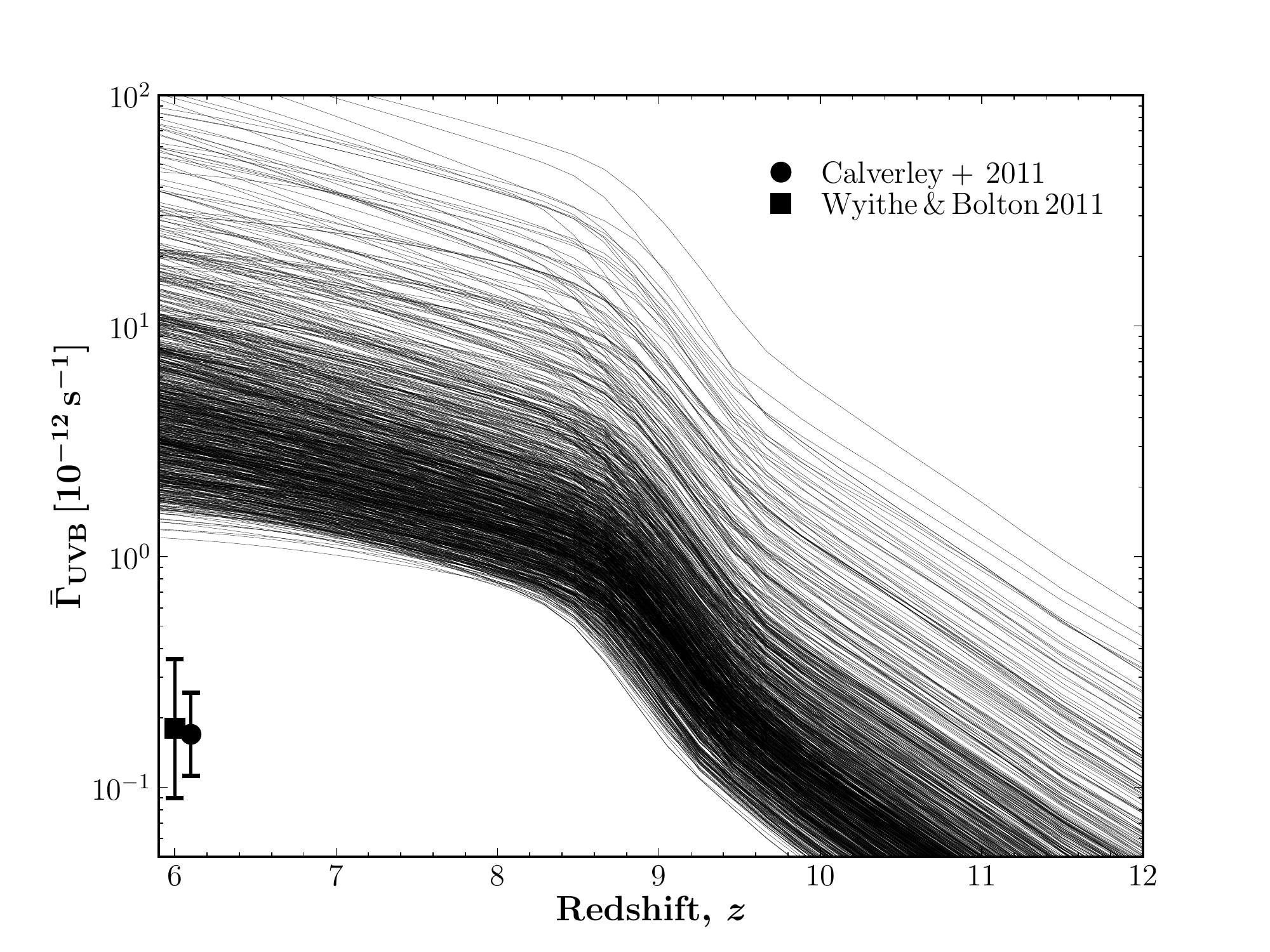}
	\end{center}
\caption[]{A comparison of the mean UV background radiation ($\bar{\Gamma}_{\rm UVB}$) from a random sample of 500 models drawn from the posterior (black curves) in excess of the LOFAR 21-cm PS upper limits against observed constraints from the proximity zones of high-$z$ quasars \citep{Calverley:2011,Wyithe:2011}.}
\label{fig:Gamma}
\end{figure}

In Figure~\ref{fig:Gamma} we compare the mean UV background photoionisation rate from 500 randomly sampled models from the posterior in excess of the LOFAR upper limits to observational constrains extracted from the proximity zones of $z>6$ QSOs \citep[e.g.][]{Calverley:2011,Wyithe:2011}. Here, the vast majority of the models ruled out by LOFAR are at least two orders of magnitude larger in amplitude than those inferred from the observational constraints ($\gtrsim3\sigma$). This is consistent with the picture of an excessive amount of ionising photons being required to ensure reionisation is $\sim50$ per cent complete by $z\approx9.1$. As a result, these models drastically overproduce the mean background photo-ionisation rate.

\subsection{Disfavoured IGM properties}

\begin{figure} 
	\begin{center}
	  \includegraphics[trim = 1.2cm 0.8cm 0cm 0cm, scale = 0.48]{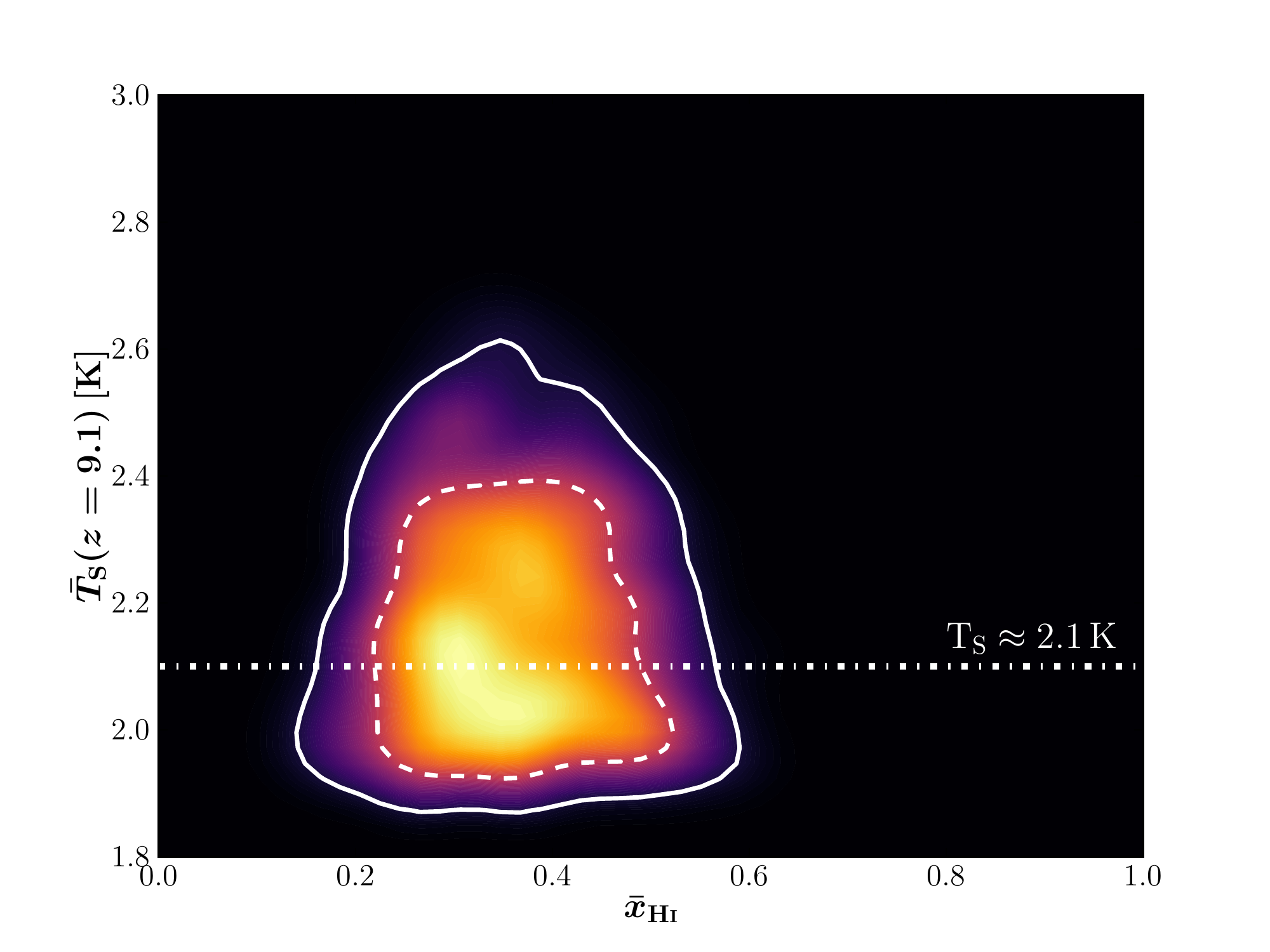}
	\end{center}
\caption[]{Two dimensional marginalised constraints on the IGM spin temperature, $T_{\rm S}$, and the IGM neutral fraction, $ \bar{x}_{\hi{}}$. The white dashed (solid) contours corresponds to the 68th (95th) percentiles while the horizontal white dot-dashed line corresponds to the value of the adiabatically cooled neutral IGM gas at mean density from \textsc{RECFAST}. }
\label{fig:TS}
\end{figure}

Finally, we directly investigate the inferred limits on the IGM spin temperature from the models found to be in excess of the LOFAR upper limits at $z\approx9.1$. Note, these are derived limits obtained by marginalising over the full posterior of the model parameters. These limits are primarily driven by those models we rule out with extremely low soft-band X-ray luminosities (i.e. ${\rm log_{10}}\left(L_{{\rm X}<2\,{\rm keV}}/{\rm SFR}\right) \lesssim 39.04$ from Section~\ref{sec:astro}). In the absence of any heating source for the IGM, the neutral hydrogen gas will adiabatically cool. At $z\approx9.1$, we infer this adiabatically cooled value for the neutral IGM gas at mean density to be $T_{\rm S} \approx2.1$~K \citep[\textsc{RECFAST};][]{Seager:1999p4330,Seager:2000p4329}.

In Figure~\ref{fig:TS} we provide the two dimensional marginalised posterior for the IGM spin temperature, $T_{\rm S}$ and the IGM neutral fraction, $\bar{x}_{\hi{}}$ after marginalising over all parameters outlined in Section~\ref{sec:fiducial}. The horizontal dot-dashed line corresponds to the value of the adiabatically cooled neutral IGM at mean density, $T_{\rm S} \approx2.1$~K. Note, within \cmfst{} it is possible to have temperatures below $T_{\rm S} \approx2.1$~K. This arises owing to adiabatic cooling and Compton heating in a non-uniform IGM. Due to non-linear structure evolution (i.e. more volume is in voids than overdensities) the volume weighted $T_{\rm S}$ can be below the limit at mean density (i.e. $T_{\rm S} \approx2.1$~K). Figure~\ref{fig:TS} implies that over a IGM neutral fraction of $0.15 \lesssim \bar{x}_{\hi{}} \lesssim 0.6$, the IGM spin temperature is disfavoured by LOFAR at 95 per cent confidence at $T_{\rm S} \lesssim2.6$~K. This implies that the IGM must have undergone, at the bare minimum, the slightest amount of heating by X-rays. 

\citet{Ghara:2020} quote their own 95 per cent disfavoured limit on the IGM spin temperature of $T_{\rm S} \lesssim2.9$~K, however, this is inferred from their uniform IGM spin temperature model (with their inhomogeneous limit instead expressed as a volume weighted fraction of heating). Nevertheless, this limit is broadly consistent with the limit from our own analysis using a non-uniform IGM spin temperature.

\section{Conclusion} \label{sec:Conclusion}

Recently, LOFAR published new upper limits on the 21-cm PS from 141 hours of data at $z\approx9.1$ \citep{Mertens:2020}. This corresponds to a factor $\sim8$ improvement over their best previous upper limit \citep{Patil:2017}. Using \cmmc{}, an MCMC sampler of 3D reionisation simulations, we explore regions of astrophysical parameter space that are inconsistent with the observed limits. We directly forward-model the 3D cosmic 21-cm signal which includes: (i) a mass-dependent star-formation rate and escape fraction to be able to self-consistently produce UV LFs, (ii) inhomogeneous recombinations, (iii) the evolution of the inhomogeneous IGM spin temperature and (iv) an on-the-fly ionising photon non-conservation corrections.

In terms of the astrophysical parameters responsible for reionisation, we find that the upper limits presented by LOFAR are able to primarily limit the X-ray luminosity of the first sources. At 95 per cent confidence, LOFAR disfavours the soft-band X-ray luminosity, $L_{{\rm X}<2\,{\rm keV}}/{\rm SFR}$ to ${\rm log_{10}}\left(L_{{\rm X}<2\,{\rm keV}}/{\rm SFR}\right) \lesssim 39.04$. These limits currently sit below expectations from local populations of star-forming galaxies \citep{Mineo:2012p6282}, stacked {\it Chandra} observations \citep{Lehmer:2016p7810} and predictions at high-redshifts by population synthesis models \citep{Fragos:2013p6529}. Nevertheless, this implies that the X-ray background must have been sufficiently large to provide some level of heating to the IGM.

For the UV galaxy properties, we find that the regions disfavoured by LOFAR are heavily tied to the observational redshift of $z\approx9.1$. That is, to be able to exceed the LOFAR upper limit, reionisation must have been $\sim50$ per cent complete to maximise the amplitude of the 21-cm PS. As such, the LOFAR upper limits disfavour the normalisations of the mass dependent star-formation rate, $f_{\ast,10}$ and escape fraction, $f_{{\rm esc},10}$, to be above $\sim3.1$ and $\sim2.3$ per cent, respectively. Finally, a minimum source mass above ${\rm log_{10}}(M_{\rm turn}) \gtrsim 8.13$ is disfavoured at 95 per cent confidence. This shows the value of having tools such as \cmmc{} to perform forward modelling of the cosmic 21-cm signal to be able to infer the astrophysics from observations of the 21-cm signal in a fully Bayesian framework. Note however, this interpretation is specific to our astrophysical parameterisation.

To highlight the advantage of using \cmmc{} we then compared the models disfavoured by the LOFAR upper limits against a range of existing observation constraints within the literature. We compared against: (i) observational constraints and limits on the IGM neutral fraction, (ii) the observed UV LFs at $z=6$, 7, 8 and 10, (iii) the electron scattering optical depth and (iv) the mean UV background photoionisation rate. In most instances the astrophysical models disfavoured by LOFAR were inconsistent with existing observational constraints by $\gtrsim2\sigma$.

Finally, we explored limits on the IGM spin temperature, $T_{\rm S}$, due to the lack of X-ray heating inferred in the models disfavoured by LOFAR. Over an IGM neutral fraction range of $0.15 \lesssim \bar{x}_{\hi{}} \lesssim 0.6$, the LOFAR upper limits imply a 95 per cent confidence limit of $T_{\rm S} \gtrsim2.6$~K. In comparison, the adiabatically cooled limit at $z\approx9.1$ is $T_{\rm S}\approx2.1$~K (at mean density), which implies the IGM must have undergone some level of X-ray heating. This interpretation is consistent with the results presented in \citet{Ghara:2020} and provides a consistency check that the recovered bulk IGM properties from the current LOFAR limits are insensitive to the details of the modelling, including the type of semi-numerical simulation, galaxy parametrisation, and use of emulators in the MCMC.

With the amplitude of the existing limits on the 21-cm PS presented by LOFAR still being relatively large, and only at a single observed frequency, this limits our ability to constrain the astrophysics of the sources responsible for reionisation. However, in the near future these limits will continue to reduce in amplitude as LOFAR processes more observational data and  improves their analysis pipeline, while at the same time also providing multiple frequency limits. Using the framework showcased here, these upcoming limits can start ruling out otherwise viable galaxy evolution models.

\section*{Acknowledgements}

Parts of this research were supported by the Australian Research Council Centre of Excellence for All Sky Astrophysics in 3 Dimensions (ASTRO 3D), through project number CE170100013. AM acknowledges funding support from the European Research Council (ERC) under the European Union's Horizon 2020 research and innovation programme (grant agreement No 638809 -- AIDA -- PI: AM). FGM and LVEK would like to acknowledge support from an SKA-NL Roadmap grant from the Dutch Ministry of Education, Culture and Science. GM is thankful for support by Swedish Research. Council grant 2016-03581. SZ acknowledges support from the Israeli Science Foundation (grant no. 255/18). ITI was supported by the Science and Technology Facilities Council [grant numbers ST/I000976/1 and ST/T000473/1] and the Southeast Physics Network (SEP-Net). 

\bibliography{Papers}

\end{document}